\documentclass[preprint,
prb]{revtex4}

\usepackage{amsmath}
\usepackage{amssymb}
\usepackage{amsthm}
\newcommand{{\bfr}}{\mbox{\boldmath$r$\unboldmath}}
\newcommand{{\bfv}}{\mbox{\boldmath$v$\unboldmath}}
\newcommand{{\bff}}{\mbox{\boldmath$f$\unboldmath}}
\newcommand{{\bfF}}{\mbox{\boldmath$F$\unboldmath}}
\newcommand{{\bfA}}{\mbox{\boldmath$A$\unboldmath}}
\newcommand{\gradv}{\boldsymbol{\nabla}}
\def\v#1{{\bf#1}}

\begin{document}
\title{Author's response}

\author{Jos\'e A. Heras}
\email{herasgomez@gmail.com}
\affiliation{Departamento de Ciencias B\'asicas, Universidad Aut\'onoma Metropolitana, Unidad Azcapotzalco, Av. San Pablo No. 180, Col. Reynosa, 02200, M\'exico D. F. M\'exico}

\maketitle

Jefimenko\cite{1} claims that my derivation of Maxwell's equations from the continuity equation\cite{2} is equivalent to his derivation of these equations presented in a recent paper.\cite{3}  However, there is a formal difference between the two derivations. The derivation given by Jefimenko is based on three postulates: (1) the time-dependent generalization of Coulomb's Law, (2) the time-dependent generalization of the Biot-Savart law, and (3) the continuity equation. He considers the generalizations of the Coulomb and Biot-Savart laws ``as the fundamental electromagnetic equations" and concludes that ``Maxwell's equations appear as derived equations." However, such generalized laws are equivalent to the familiar retarded electric and magnetic fields [Eqs.~(12) and (13) of Ref.~3]. Jefimenko's derivation explicitly  uses this familiar form of the retarded fields to derive the Maxwell equations. Jefimenko's paper\cite{3} derives an expected result: Maxwell's equations can be obtained by postulating a retarded form of their solutions. 

In contrast, my derivation of Maxwell's equations is based on an existence theorem: Given localized sources $\rho$ and $\v J$ satisfying the continuity equation $\gradv\cdot\v J+\partial\rho/\partial t=0,$ there exist retarded fields $\v F$ and $\v G$ that satisfy the field equations 
$\gradv\cdot\v F=\alpha\rho,\;\gradv\cdot\v G= 0,\gradv\times \v F+\gamma\partial \v G/\partial t= 0$ and 
$\gradv\times \v G-(\beta/\alpha)\partial \v F/\partial t=\beta\v J,$ where the arbitrary positive constants $\alpha$, $\beta$, $\gamma$, and $c$ are related by $\beta\gamma c^2=\alpha $. Crucial for obtaining this theorem are Eqs.~(18) and (20) of Ref.~2, which are implied by the continuity equation. These equations show the formal existence of the fields $\v F$ and $\v G$ defined by retarded values of the sources $\rho$ and $\v J$ [Eqs.~(2) of Ref.~2]. When $\rho$ and $\v J$ are identified with the usual electromagnetic charge and current densities and appropriate choice of constants made, the fields $\v F$ and $\v G$ become the generalized Coulomb and Biot-Savart laws, $\v F=\v E$ and $\v G=\v B$, and their associated field equations become Maxwell's equations. My paper\cite{2} proves that Maxwell's equations and the generalized Coulomb and Biot-Savart laws can formally be obtained from the continuity equation for the electric charge and current densities. The formal difference between Jefimenko's derivation of Maxwell's equations and  mine is that the former uses three postulates while the latter only one. As far as an axiomatic presentation of Maxwell's equations is concerned, the number of required postulates becomes important.

It can be argued that my derivation of Maxwell's equations involves other implicit physical assumptions like retardation (causality) and the propagation at speed of light $c.$ However, in the formulation of the existence theorem the parameter $c$ plays the role of an arbitrary constant, so choosing $c$ to be the speed of light in a vacuum is not a new postulate, but only a special value. Analogously, in my derivation of Maxwell's equations I postulated the validity of the continuity equation at all times, so evaluating it at a particular time (the retarded time) is not a new postulate, but only a special case of the first. 

{}
,

\end{document}